\documentclass[aps, 12pt, showkeys, nofootinbib]{revtex4}

\newcommand{\be}{\begin{eqnarray}}
\newcommand{\ee}{\end{eqnarray}}
\newcommand{\ba}{\begin{array}}
\newcommand{\ea}{\end{array}}
\newcommand{\ds}{\displaystyle}

\newcommand{\nn}{\nonumber}
\newcommand{\pa}[1]{\left(#1\right)}
\newcommand{\paq}[1]{\left[#1\right]}

\newcommand{\K}{\mathbf{k}}
\newcommand{\Pp}{\mathbf{p}}

\newcommand{\Q}{{\mathbf q}}
\newcommand{\pp}{{\mathbf p}}

\newcommand{\ta}{{\tt a}}

\usepackage[english]{babel}
\usepackage{amsmath}
\usepackage{amssymb}
\usepackage{graphicx,color}
\makeatletter\AtBeginDocument{\let\@elt\relax}\makeatother

\usepackage[normalem]{ulem}

\begin{document}

\title{Gravitational Multipole Renormalization}

\author{Gabriel Luz Almeida}
\affiliation{Departamento de F\'\i sica Te\'orica e Experimental, Universidade Federal do Rio Grande do Norte, Avenida Senador Salgado Filho, Natal-RN 59078-970, Brazil}

\author{Stefano Foffa}
\affiliation{D\'epartement de Physique Th\'eorique and Center for Astroparticle Physics, Universit\'e de Gen\`eve, CH-1211 Geneva, Switzerland}

\author{Riccardo Sturani}
\affiliation{International Institute of Physics, Universidade Federal do Rio Grande do Norte, Campus Universit\'ario, Lagoa Nova CP:1613, Natal-RN 59078-970, Brazil}

\email{gabriel.luz@fisica.ufrn.br, stefano.foffa@unige.ch, riccardo.sturani@ufrn.br}

\begin{abstract}
We study the effect of scattering gravitational radiation off the
    static background curvature, up to second order in Newton constant,
    known in the literature as tail and tail-of-tail processes,
    for generic electric and magnetic multipoles.
  Starting from the multipole expansion of composite compact objects, and
  as expected due to the known electric quadrupole case,
  both long- and short-distance (UV) divergences are encountered.
  The former disappear from properly defined observables, the latter are
  renormalized, and their associated logarithms give rise to a classical
  renormalization group flow.
  UV divergences alert for incompleteness of the multipolar
  description of the composite source and are expected not to be present
  in a UV-complete theory, as explicitly derived in the literature for the case of
  conservative dynamics.
  Logarithmic terms from tail-of-tail processes associated to generic magnetic
  multipoles are computed in this work for the first time.
\end{abstract}

\keywords{Classical General Relativity, Gravitational Radiation, Post-Newtonian expansion}

\maketitle
\nopagebreak

\section{Introduction}
\label{sec:intro}
The recent detections of gravitational waves emitted by compact binary coalescences
\cite{LIGOScientific:2018mvr,Abbott:2020niy,LIGOScientific:2021qlt}, observed by the LIGO \cite{TheLIGOScientific:2014jea} and Virgo
\cite{TheVirgo:2014hva} large interferometric detectors,
made the compelling case for improving the knowledge of binary system
dynamics, as its features are imprinted in the details of the detected
waveforms.

The starting point of this work is the multipolar action, describing the
coupling of a compact source to an external gravitational field in General
Relativity.
When the multipoles describe a composite source with internal velocity $v$ and
size $r$, like in the case of compact binary coalescence, the multipolar
expansion parameter is $v$, and in this case the gravitational radiation emitted
by a time-varying multipole has angular frequency $\omega \sim v/r$.

Building on the multipole expansion, we study a specific class of post-Minkowskian (PM)
corrections up to second order in the Newton constant $G_N$.
At $O(G_N)$ beyond leading-order emission, one encounters leading nonlinear \emph{hereditary}
effects, i.e., terms depending on the history of the source rather than on an instantaneous state
at retarded time.
Historically, these  have been divided into \emph{memory} and \emph{tail} effects \cite{Blanchet:1992br},
the former arising from scattering of radiation onto radiation \cite{Christodoulou:1991cr},
the latter from scattering of radiation onto the static background curvature sourced
by the total mass $E$ of the system \cite{Blanchet:1987wq}.
The denominations are related to the nature of the phenomenological effects they have on the waveform: The tail part of the waveform
arrives later than the ``wave front,'' being delayed by the scattering, and then smoothly fades off with time;
the memory part is a persistent zero-frequency effect which is still present well after the wave front has passed.

While hereditary in the waveform, radiation-radiation scattering leads to a vanishing effect
in the emitted flux \cite{Blanchet:2013haa}
and to an instantaneous (i.e.,~nonhereditary) contribution to the conservative energy \cite{Foffa:2019eeb};
tail effects, on the other hand, give a hereditary contribution to the waveform \cite{Blanchet:1987wq} and to the
conservative energy \cite{Foffa:2011np} (later confirmed in Ref. \cite{Galley:2015kus})
while giving an instantaneous contribution to the flux emission from circular orbits
\cite{Goldberger:2009qd}.
The scattering of radiation off the angular-momentum-dependent static background curvature leads
to instantaneous terms both in the waveform \cite{Faye:2012we} and in the conservative energy shift
\cite{Foffa:2019eeb} and no contribution to the flux.

In particular, only the (mass) tail-corrected emission process involves a large-distance, or infrared (IR),
divergence, as thoroughly explained in Ref. \cite{Goldberger:2009qd}, which, however, disappears
from suitably defined observables.
In the waveform, the IR tail divergences are relatively imaginary with respect to the leading order, and
they exponentiate to a pure phase, so disappearing from the flux.
While, in principle, still showing up in the waveform, analogous to the
well-known infinite phase shift induced by the Coulomb potential in scattering amplitudes \cite{Weinberg:1965nx},
one has to consider that actual detections do not measure the instantaneous absolute value of the phase
but phase differences between different times, and the infinity cancels out of any observable
quantity \cite{Porto:2012as}.
Note, however, that finite contributions of the tail effect for different
multipoles are different, and their nonzero difference \emph{is} physical,
while the IR divergent part is common to all multipoles \cite{Faye:2014fra}
and cancels out in the difference.

Note that observability of the finite shift in the waveform phase generated by the
tail effect has already been investigated long ago in Refs. \cite{Blanchet:1994ex,Blanchet:1994ez}, and, unfortunately, the possibility of it being measured is scarce,
as such an effect appears as $G_NE\omega\sim v^3$ correction to the leading-order 
phase which goes as $v^{-5}$, hence a fourth-order post-Newtonian (PN)
  effect \cite{Blanchet:1994ez},
where $v^2\sim G_NE/r$ is the expansion parameter of the PN approximation.
Current knowledge of PN-expanded waveforms stops at 3.5PN order;
see Ref. \cite{Blanchet:2013haa} for a review and Ref. \cite{LIGOScientific:2020ros} for
the most recent tests on real data.
Note that finite contributions of the tail affect the waveform phase
at the same order as a shift $\Delta t$ in the arrival time of the signal,
which enters the phase with a term
$\sim 2\pi f\Delta t\sim v^3 (\Delta t/G_NE)$.

The main focus of the present work is the analysis of (mass) tail-of-tail effects
at waveform level or, equivalently, in the language of field theory, in one-point 
amplitudes. They come with both IR and UV divergences; the former
are consistent with the exponentiation to a phase of the simple tail IR
divergences, and the latter have associated logarithmic terms that give rise to
renormalization group equations, which can be integrated to compute all-orders
leading logarithmic corrections, as already done for the logarithms from the
electric quadrupole case \cite{Blanchet:2019rjs}.

In particular, we generalize the computation of logarithmic
terms in tail-of-tail processes, already known in the electric case from the
results obtained in Ref. \cite{Anderson:1982fk} for the mass quadrupole and in
Ref. \cite{Blanchet:1987wq} for all the electric multipoles,
to magnetic multipoles at all orders.
While sharing the same topology, diagrams of increasing multipole order
become more intricate because of the presence of an increasing
number of momenta. In PN scaling, moving from a multipole to the following one
adds a power of $v$ to the coupling; hence, tail diagrams involving the
electric (magnetic) $2^{n}$-multipole
affect one-point amplitudes starting at $1/2+n/2$ ($1+n/2$) PN order.
Multipoles corrected by gravitational self-interactions are also called in the
literature \emph{radiative} multipoles \cite{Blanchet:2013haa}, to differentiate
from \emph{source} multipoles, which instead designate the source terms in the
fundamental multipolar expansions.

Note, however, that, when multipoles of composite objects like binary systems
are expressed in terms of individual binary constituents, they can naturally
be expanded in $v^2$, i.e., in a PN series, whose
terms are determined by a \emph{matching} procedure, which for the mass quadrupole has been
completed in an effective field theory (EFT) framework up to second PN order
\cite{Leibovich:2019cxo} and to fourth PN order in the multipolar-post-Minkowskian approach \cite{Marchand:2020fpt}.

By analogy with the conservative dynamics case treated in detail in Ref. \cite{Foffa:2019yfl},
we expect that the UV divergence in the tail-of-tail process will be canceled
by analogous divergences in the expression of the PN-corrected source multipoles,
to leave a finite, consistent result.
After all, the multipole expansion is bound to fail at a short enough distance,
i.e.,~when the actual internal structure of the composite system becomes important.

The paper is structured as follows: In Sec.~\ref{sec:method}, we give an overview
of the method, treating in detail the known case of tail process,
building on which we obtain new results for the tail-of-tail process
in Sec.~\ref{sec:res}. Section~\ref{sec:sdisc} concludes the present work with
a discussion of the results.

\section{Method}
\label{sec:method}

\subsection{Generalities}
We will proceed from and expand along the lines of Ref. \cite{Goldberger:2009qd}, which
applies to the radiative gravitational sector the EFT approach developed in Ref. \cite{Goldberger:2004jt}, known as nonrelativistic General Relativity.

At a large distance from the source, its interaction with gravity can be encoded
in terms of multipoles as in the following effective Lagrangian, whose form is
uniquely dictated by the symmetries and scaling of the theory:\footnote{We use the mostly plus metric signature and the speed of light $c=1$ throughout the paper. Latin indices run over $\{1,2,3\}$ and are raised and lowered by Kronecker deltas.}
\begin{align}
\label{eq:smult}
{\cal S}_{\text{mult}} &=\int {\rm d}t \left(\frac 12 Eh_{00}-\frac 12\epsilon^{ijk}L_ih_{0j,k}
     -\frac 12I^{ij}{\cal E}_{ij}-\frac 16 I^{ijk}{\cal E}_{ij,k}+
     \frac 23 J^{ij}{\cal B}_{ij}+\ldots\right)\nonumber\\
     &=\int {\rm d}t\left[\frac 12 Eh_{00}-\frac 12\epsilon^{ijk}L_ih_{0j,k} 
     -\sum_{r\geq 0}\left(c^{(I)}_r I^{iji_1\cdots i_r}\partial_{i_1}\cdots\partial_{i_r}{\cal E}_{ij}-c^{(J)}_r J^{iji_1\cdots i_r}\partial_{i_1}\cdots\partial_{i_r}{\cal B}_{ij}\right)
\right]\,,
\end{align}
with \cite{Ross:2012fc}
\be
c_r^{(I)}=\ds\frac 1{\pa{r+2}!}\,,\quad c_r^{(J)}=\ds\frac {2\pa{r+2}}{\pa{r+3}!}\,,
\ee
where $E$ and $L_i$ are, respectively, energy and angular momentum, $I^{iji_1\cdots i_r}$
($J^{iji_1\cdots i_r}$) are generic electric (magnetic) source $2^n$-poles
for $n\geq 2$, $n=r+2$ (i.e., from quadrupole on), and ${\cal E}_{ij}$ and ${\cal B}_{ij}$
denote, respectively, the electric and magnetic part of the Riemann tensor.
\footnote{Terms proportional to the center of mass position and velocity in the multipole
expansions have been neglected. We denote by an over-dot the time
derivative and by $\epsilon_{ijk}$ the 3-dimensional Levi-Civita tensor.
If the $d$-dimensional Levi-Civita tensor is used instead, one has $\epsilon_{ijk}\epsilon_{ilm}=\pa{d-2}\pa{\delta_{jl}\delta_{km}-\delta_{jm}\delta_{kl}}$ and the extra $d-2$ factor must be compensated by
an inverse rescaling of the magnetic multipoles $J^{iji_1\cdots i_r}$.}

In case one is interested in applications to compact binary systems, the source
multipoles appearing in Eq. (\ref{eq:smult}) can be explicitly related
to individual constituents' parameters by means of a matching procedure,
as done up to 2PN for the mass quadrupole $I^{ij}$ within the EFT approach in
Ref. \cite{Leibovich:2019cxo} and to higher orders within the multipolar
Minkowskian formalism; see \cite{Blanchet:2005tk,Faye:2014fra,Blanchet:2008je,Marchand:2020fpt,Henry:2021cek}, and references therein.

In the present work, we are mainly interested in the  universal properties
(i.e., not depending on the short-scale features of the source) of the
gravitational waveform, so our focus will not be on the matching procedure
but rather on the study of emission amplitudes, as expressed in terms of the
generic multipoles $I^{iji_1\cdots i_r}$ and $J^{iji_1\cdots i_r}$, with particular
emphasis on the divergences appearing in dimensional regularization and on the
associated logarithmic terms.
We are also not studying here conservative effects associated to emission
  and reabsorption of radiative modes, for which we refer to Refs.
  \cite{Foffa:2019eeb,Almeida:2020mrg}.

We work in the harmonic gauge, as in Refs. \cite{Blanchet:2013haa,Bernard:2015njp},
which is equivalent to using the following form for the pure (bulk) gravity action:
\be\label{az_bulk}
S_{bulk}= 2 \Lambda^2 \int {\rm d}^{d+1}x\sqrt{-g}\left[ R(g)-\frac{1}{2}\Gamma_\mu\Gamma^\mu\right]\,,
\ee
where $R(g)$ is the Ricci scalar, $\Gamma^\mu \equiv g^{\rho\sigma}\Gamma^\mu_{\rho\sigma}$, $\Gamma^\mu_{\rho\sigma}$ being the standard Christoffel
coefficients,
and $\Lambda^{-2}\equiv 32 \pi G_N \mu^{3-d}$. Note that for
the number of purely spatial dimensions $d\neq 3$ an
inverse length $\mu$ appears, as it is necessary to relate $\Lambda$,
which has dimensions $(\mathrm{mass}/\mathrm{length}^{d-2})^{1/2}$, to the ordinary 3+1-dimensional
Newton constant $G_N$.

We find it useful to decompose the metric via a Kaluza-Klein parameterization \cite{Kol:2007bc}:
\be
\label{met_nr}
g_{\mu\nu}=e^{2\phi/\Lambda}\pa{
\ba{cc}
-1 & A_j/\Lambda \\
A_i/\Lambda &\quad e^{- c_d\phi/\Lambda}\gamma_{ij}-
A_iA_j/\Lambda^2\\
\ea
}\,,
\ee
with $\gamma_{ij}\equiv\delta_{ij}+\sigma_{ij}/\Lambda$,
and $c_d\equiv 2\frac{(d-1)}{(d-2)}$.
In this decomposition, one can write at linear order
\be
\ba{rcl}
\ds\Lambda{\mathcal E}_{ij}&\simeq&\ds-\frac 12\pa{\ddot\sigma_{ij}-\dot A_{i,j}-\dot A_{j,i}}
+\phi_{,ij}+\frac{\delta_{ij}}{d-2}\ddot\phi+O(h^2)\\
\ds\Lambda{\mathcal B}_{ij}&\simeq&\ds\frac 14\epsilon_{ikl}\paq{\dot\sigma_{jk,l}-\dot\sigma_{jl,k}+A_{l,jk}-A_{k,jl}+\frac 2{d-2}\pa{\dot\phi_{,k}\delta_{jl}-
\dot\phi_{,l}\delta_{jk}}}+O(h^2)\,,
\ea
\ee
where $h$ denotes the generic metric perturbation around Minkowski spacetime.

\begin{figure}
  \begin{center}
    \includegraphics[width=.20\linewidth]{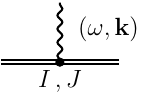}
  \caption{Feynman diagram representing the leading-order emission amplitude.}
  \label{fig:emLO}
  \end{center}
\end{figure}

The radiative, transverse-traceless part of the metric
perturbation corresponds to the transverse-traceless part of $\sigma_{ij}$
(also denoted $\sigma_{ij}$ for simplicity), and the leading-order amplitude
for emission of gravitational mode with on-shell 4-momentum $(\omega,\K)$,
with $\omega^2=\K^2$, by a generic electric ($I$) or magnetic ($J$) multipole
can be written as
\footnote{Our choice for the metric signature implies that uppercase spatial indices are equivalent to lowercase ones. Taking advantage of this fact, we will allow a little abuse of notation in indices position to make
      equations more appeasing to the eye.}
\begin{equation}
\label{eq:emLO}
i \mathcal{A}_0(\omega,\K)=
\sum_r\frac{(-i)^{r+1}}{2\Lambda}\sigma_{ij}^{*}(\omega,\K) k_{i_1}\cdots k_{i_r}\left[
{c_r}^{(I)} \omega^2 I^{iji_1\cdots i_r}(\omega)
+
c_r^{(J)}\omega\epsilon_{ikl} k_l J^{jki_1 \cdots i_r}(\omega)
\right]\,,
\end{equation}
with its corresponding Feynman diagrams in Fig.~\ref{fig:emLO}.

By applying standard tools for Feynman diagram computations, one can derive
$O(G_N)$ and $O(G_N^2)$ corrections to the emission amplitude in Eq.~(\ref{eq:emLO}),
which will be shown in the next sections.
The explicit expression for propagators and interaction vertices can be read
from Ref. \cite{Foffa:2019rdf} and will not be reported here, with the only
modification that for emission processes retarded Green's functions have to be
used, which can be represented as
\be
\label{eq:rgf}
G_R(\omega,\K)=\lim_{\ta\to 0^+}\frac 1{\pa{\omega+i\ta}^2-\K^2}\,,
\ee
and in all propagators in the rest of this paper we will denote by $\ta$ an
arbitrary small, positive quantity.
The gravitational field can be obtained (at leading order) in Fourier space
  by multiplying the (leading-order) amplitude (\ref{eq:emLO}) by the retarded
  Green's function (\ref{eq:rgf}), as it is causally determined by the source.
  Boundary conditions are specified by the pole displacement in the inverse
  space representation of the Green's function; hence, their effect shows up
  only for the region of momenta having $|\K|=\omega$. \footnote{Note
    that in Ref. \cite{Goldberger:2009qd} Feynman Green's functions
    have been adopted instead. As pointed out in Ref. \cite{Galley:2015kus}, such a
    prescription does not generally allow one to obtain the correct imaginary part
    of the amplitude (see also footnote 5).}

\subsection{Tails}
\label{ssec:met_tail}
\begin{figure}
\begin{center}
\includegraphics[width=.30\linewidth]{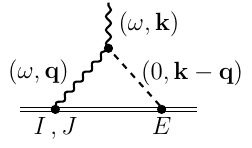}
\caption{Feynman diagram representing the tail emission amplitude.}
\label{fig:tail}
\end{center}
\end{figure}

The computation of the tail amplitude involving the energy and the electric
quadrupole was first derived in Ref. \cite{Blanchet:1987wq}, and it has been rederived
in Ref. \cite{Goldberger:2009qd} with effective field theory methods;
here, we report the results involving generic electric and magnetic multipoles, as represented in Fig.~\ref{fig:tail}
as a warm-up for subsequent calculations. Note that the gravitational mode
attached to the conserved energy $E$ has a vanishing time component.

Adopting the notation $\int_\Q\equiv \int \frac{{\rm d}^dq}{\pa{2\pi}^d}$,
in the electric case one has ($\omega^2=\K^2$)
\begin{align}
\label{eq:tailE}
i{\cal A}^{(e)}_{r-tail}(\omega,\K) = &(-i)^{r+1} \left(\frac{E c_r^{(I)}}{4\Lambda^3}\right)I^{iji_1\cdots i_r}(\omega)
\int_{\Q}\frac 1{\paq{\Q^2-\pa{\omega+i\ta}^2}}\frac 1{\pa{\K-\Q}^2}\times
q_{i_1}\cdots q_{i_r}\nonumber\\
&\times\paq{\omega^4\delta_{ai}\delta_{bj}+2\omega^2q_i\pa{k-q}_a\delta_{bj}
+\frac {2}{c_d}q_iq_j\pa{k-q}_a\pa{k-q}_b} \sigma_{ab}^{*}(\omega,\K)\nn\\
  \simeq& i{\cal A}_{r0}^{(e)}(\omega,\K)\pa{i G_N E \omega}
  \paq{-\frac{\pa{\omega+i\ta}^2}{\tilde{\mu}^2}}^{\epsilon_{\mathrm{IR}}/2}
        \paq{\frac{2}{\epsilon_{\mathrm{IR}}}-2\kappa_{r+2}+O(\epsilon_{\mathrm{IR}})}\,,
\end{align}
where ${\cal A}_{r0}^{(e)}$ is the electric part of the $2^{2+r}$-multipole
in Eq.~(\ref{eq:emLO}),
$\epsilon\equiv d-3$, $\tilde{\mu}^2\equiv \pi\mu^2e^{-\gamma}$, with $\gamma$ the
Euler constant,
\be
\kappa_{r+2}\equiv\frac{2r^2+13 r+22}{(r+2)(r+3)(r+4)}+H_r\,,
\ee
and $H_r$ is the $r$th \emph{harmonic number} defined by $H_r\equiv \sum_{i=1}^r 1/i$.
The second line in Eq.~(\ref{eq:tailE}) is determined by the bulk
  interactions of the tail diagram, which depends on $\sigma^2\phi$,
  $\sigma A\phi$, and $\sigma\phi^2$ interactions contained in the Einstein-Hilbert action.
Expanding also the factor $\paq{-\pa{\omega+i\ta}^2/\tilde\mu^2}^{\epsilon/2}$ in
Eq. (\ref{eq:tailE}) for $\epsilon\to 0$, recalling the cut in the negative real
semiaxis of the $\omega$ complex plane, one finally gets
\footnote{Note the
    presence of the ${\rm sgn}(\omega)$ term in Eq.~(\ref{eq:tailE_bis}), which
    is necessary to ensure that the tail corrections satisfy the reality property
    $\mathcal{A}^*(\omega)=\mathcal{A}(-\omega)$, to ensure
  a real waveform in direct space. Had one used Feynman Green's function, one would have had
  $(\omega^2+i\ta)$ replacing $\pa{\omega+i\ta}^2$ in Eq.~(\ref{eq:tailE}),
  then obtaining $-i\pi$ instead of $-i\pi{\rm sgn}(\omega)$ in
  Eq.~(\ref{eq:tailE_bis}).}
\be
\label{eq:tailE_bis}
i{\cal A}^{(e)}_{r-tail}(\omega,\K) \simeq i {\cal A}_{r0}^{(e)}(\omega,\K)
\pa{i G_N E \omega}\paq{\frac{2}{\epsilon_{\mathrm{IR}}}-2\kappa_{r+2}-i\pi\, {\rm sgn}(\omega)+
\log\pa{\frac{\omega^2}{\tilde\mu^2}}}\,.
\ee

An analogous calculation for the magnetic multipole gives
\begin{align}
\label{eq:tailM}
i{\cal A}^{(m)}_{r-tail}(\omega,\K) =& (-i)^{r+1} \left(\frac{E c_r^{(J)}}{4\Lambda^3}\right)\omega\epsilon_{ikl} J^{jki_1\cdots i_r}(\omega)
\int_{\Q}\frac 1{[\Q^2-\pa{\omega+i\ta}^2]}\frac 1{\pa{\K-\Q}^2}\times
q_{i_1}\cdots q_{i_r}\nn\\
&\times q_l\paq{\omega^2\delta_{aj}+q_j\pa{k-q}_a}\sigma_{ai}^{*}(\omega,\K)\nn\\
 \simeq& i{\cal A}_{r0}^{(m)}(\omega,\K)\pa{i G_N E\omega}
\paq{\frac 2{\epsilon_{\mathrm{IR}}}-2\pi_{r+2}-i\pi\, {\rm sgn}(\omega)+
\log\pa{\frac{\omega^2}{\tilde\mu^2}}}\,,
\end{align}
with
\be
\pi_{r+2}\equiv\frac{r+1}{(r+2)(r+3)}+H_{r+1}\,.
\ee
The integrals have been computed using the formulas reported in the Appendix,
and the divergences encountered here are of the IR type, hence the index
``IR'' to $\epsilon$ in Eqs.~(\ref{eq:tailE}), (\ref{eq:tailE_bis}), and (\ref{eq:tailM}). They are the leading order of an
unobservable divergent phase term common to all multipoles;
the finite terms proportional to $\kappa_{r+2},\pi_{r+2}$ (first computed in Ref. \cite{Blanchet:1995fr}) are also exponentiated to a phase \cite{Porto:2012as},
which is, however, multipole dependent and so, in principle, observable.
Note that the contribution of the $-i\pi\, {\rm sgn}(\omega)$ term in the
square brackets is real relative to ${\cal A}_0$; hence, it is the only contribution from
the tail process to the emission flux at $G_NE\omega\sim v^3$ order.

The amplitudes (\ref{eq:tailE}) and (\ref{eq:tailM}) are proportional to waveforms;
hence, they can be inverse-Fourier transformed to give the waveforms in the time
domain, with the result that the logarithmic terms
in $\omega$ are responsible for nonlocal terms in direct space (i.e., in time)
first individuated in Ref. \cite{Blanchet:1987wq}.
Note that the IR divergence arises from the loop integral displayed in
Eq.~(\ref{eq:tailE}), as it is clearly shown by changing the integration
variable to $\Q'\equiv\Q-\K$:
\be
\label{eq:tail_qual}
\left.{\cal A}_{tail}\right|_{\rm IR-div}(\omega)\propto \int_{\Q'}\frac 1{\pa{2\K\cdot \Q'+{\Q'}^2}\Q'^2}\,,
\ee
and it is present only for terms whose numerator, which is set to unity
for clarity in Eq.~(\ref{eq:tail_qual}), is nonvanishing for $\Q'\to 0$.
An analog process can be considered by replacing the energy $E$ insertion of the tail
diagram with the angular momentum $L$, which, however, comes with one gradient, i.e., one power of
$\Q'$ [see Eq.~(\ref{eq:smult})], thus having no divergence and producing a local
result both in Fourier and in direct space, as can be explicitly checked in Ref. \cite{Blanchet:2008je};
for this reason, it has been dubbed ``failed'' angular momentum tail in Ref.
\cite{Foffa:2019eeb}.

Another qualitatively different process, the \emph{memory}, can be considered
at $O(G_N)$ order. It can be obtained by replacing the conserved quantity source insertion of
the tail diagram ($E$ or $L$) with a time-dependent multipole $I'$ or $J'$,
giving rise to an amplitude of the type
\be
\mathcal{A}_{memory}(\omega)\propto\int\frac{d\omega'}{2\pi}\int_\Q
\frac{I(\omega-\omega')I'(\omega')}{\paq{\Q^2-\pa{\omega-\omega'+i\ta}^2}\paq{\pa{\K-\Q}^2-\pa{\omega'+i\ta}^2}}\,,
\ee
which is not divergent but gives rise to a product of (Fourier transformed)
dynamical multipoles, which in direct space involve a convolution in time
\cite{Blanchet:2008je}.
In particular, the contribution from
$I(\omega-\omega')I'(\omega')$ for $\omega\to 0$ gives rise to a
nonvanishing zero-frequency effect, the memory effect \cite{Christodoulou:1991cr}. 

\section{Results for the tail-of-tail process}
\label{sec:res}

We derive in this section the divergent and logarithmic parts of the more challenging tail-of-tail contributions,
at second order in $G_N E \omega$ beyond leading order
(equivalent to relative 3PN for binary systems),
which is where UV divergences make their first appearance.

The tail-of-tail contribution to the radiative multipole has been derived in detail
in Ref. \cite{Blanchet:1997jj} and in Ref. \cite{Goldberger:2009qd} for the electric quadrupole case only (terms $E^2\times I_{ij}$) within EFT methods, which we generalize in this section to the
$E^2\times(I,J)$ case, for electric and magnetic multipoles of any order.

\begin{figure}
\begin{center}
\includegraphics[width=.27\linewidth]{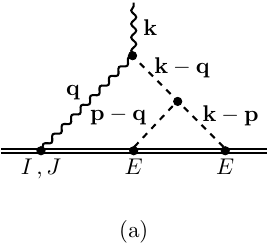}
\hspace{0.7cm}
\includegraphics[width=.27\linewidth]{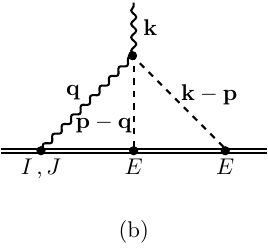}
\hspace{0.7cm}
\includegraphics[width=.27\linewidth]{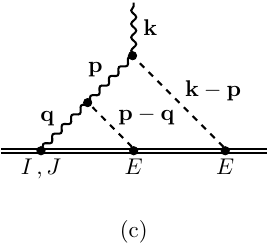}
\end{center}
\caption{Feynman diagrams describing the tail-of-tail process. We label explicitly in the figure only the space components of the momenta, the
time component being $\omega$, with $\omega^2=\K^2$ for wavy lines and
vanishing for dashed straight lines.}
\label{fig:tail2}
\end{figure}

The tail-of-tail process receives contributions from three different diagrams given in
Fig.~\ref{fig:tail2}.

The diagrams in Figs. 3(a) and 3(b)
can be computed using standard integration techniques,
bringing pure UV divergences for any multipole, as described in Ref. \cite{Goldberger:2009qd} for the quadrupole case, as can be shown as follows.
After the first loop integration over $\pp$, which can be performed via the
first equation in (\ref{eq:app_std}), and after dropping the tensor
structure for clarity, one is left
with an integral similar to the tail one of Eq. (\ref{eq:tail_qual}):
\be
\left.{\cal A}_{a,b-tail^2}\right|_{div}\sim\int_\Q\frac 1{\paq{\Q^2-\pa{\omega+i\ta}^2}\paq{\pa{\K-\Q}^2}^{m-d/2}}=
\int_{\Q'}\frac 1{\paq{2\K\cdot{\Q'}+{\Q'}^2}\paq{{\Q'}^2}^{m-d/2}}\,,
\ee
which, however, has the crucial difference from Eq. (\ref{eq:tail_qual}) of having
$m$ a positive integer, giving a half-integer exponent for the ${\Q'}^2$ term,
hence leading to a pure UV divergence, when combined with the ${\Q'}^2$ part of
the $(2\K\cdot\Q'+{\Q'}^2)$ propagator, and no IR divergence.
As noted in Ref. \cite{Goldberger:2009qd}, such diagrams correspond to the scattering
of the emitted radiation with the $1/r^2$ relativistic correction to the static
potential.

The diagrams in Figs. 3(a) and 3(b) give for the electric and magnetic case (see the Appendix for details)
\be
\label{eq:tail2_em_ab}
i {\mathcal A}^{(e,m)}_{a,b-tail^2}(\omega\,,\K) &\simeq&\ds i \mathcal{A}_{r0}^{(e,m)}(\omega\,,\K)\pa{G_N E\ \omega}^2\paq{-\frac{\pa{\omega+i\ta}^2}{\tilde{\mu}^2}}^{\epsilon_{\mathrm{UV}}}\paq{\frac{\alpha^{(e,m)}_{a,b}(r)}{\epsilon_{\mathrm{UV}}}+O(\epsilon^0)}\,,\\
\ds\alpha^{(e)}_a(r)&\equiv&\ds\frac{2r^3+3r^2-r+1}{(2 r-1) (2 r+1) (2 r+3) (2 r+5)}\,,\\
\ds\alpha^{(e)}_b(r)&\equiv&\ds -2\frac{\left(16 r^3+56 r^2+24 r -31\right)}{(2 r-1)
   (2 r+1) (2r +3) (2 r+5)}\,,\\
\alpha^{(m)}_a(r)&\equiv&\ds\frac{2r^3+11r^2+21r+17}{(2 r+1) (2 r+3) (2 r+5) (2 r+7)}\,,\\
\alpha^{(m)}_b(r)&\equiv&\ds -2\frac{\left(16 r^3+104 r^2+187 r +74\right)}{(2 r+1)
(2 r+3) (2r +5) (2 r+7) }\,.
\ee

For the more intricate diagram in Fig. 3(c), which can be decomposed in terms of the same
master integrals (\ref{eq:app_std}), we report its amplitude before integration,
split in terms of the gravitational polarization propagating in the internal wavy lines of the diagram in Fig. 3(c) ($\omega^2=\K^2$).
For the electric case, one has
\be
\label{eq:c_e}
\ba{rccll}
\ds i{\cal A}^{(e)}_c(\omega\,,\K) &=&\ds (-i)^{r+1}&\ds
\pa{-\frac{E^2 c_r^{(I)}}{4\Lambda^5}}\omega^2 I^{iji_1\cdots i_r}(\omega)&\\
&&\ds\times\int_{\pp, \Q}&\ds\frac {q_{i_1}\cdots q_{i_r}}{\paq{{\Q}^2-\pa{\omega+i\ta}^2}
\paq{{\pp}^2-\pa{\omega+i\ta}^2}(\pp-\Q)^2 (\pp-\K)^2}&\\
&&\ds\times \ds\sigma^*_{ab}(\omega,\K)&\ds
\Bigg\{-\frac{1}{2}\omega^4\delta_{ia}\delta_{jb}& \{\sigma^2\}\\
&&&\ds\quad+\omega^2\paq{q_aq_j-2p_a q_j +p_a p_j}\delta_{ib}&\{A\sigma\}\\
&&&\ds\quad -\frac{1}{c_d}q_i q_j p_a p_b &\{\phi^2\}\\
&&&\ds\quad+\frac{1}{c_d}q_i [q_j q_a -2 q_j p_a +p_j p_a]p_b&\{\phi A\}\\
&&&\ds\quad+\frac{1}{c_d}[(q-p)_i p_j p_a p_b-q_i q_j (q-p)_a q_b]&
\{\phi\sigma\}\\
&&&\ds\quad+q_i \paq{q_j p_b-q_bp_j + (\pp\cdot\Q)\delta_{bj}}p_a&\{A^2\}
\Bigg\}\\
&\simeq&\ds i \mathcal{A}_{r0}^{(e)}(\omega\,,\K)&\ds\pa{G_N E \omega}^2
\paq{-\frac{\pa{\omega+i\ta}^2}{\tilde{\mu}^2}}^{\epsilon}\paq{-\frac2{ \epsilon^2_{\rm IR}}+\frac{\alpha^{(e)}_c(r)}{\epsilon}}&\,,\\
\ea\\
\label{eq:cc_e}
\alpha^{(e)}_c(r)\equiv 2\paq{(r+1)\frac{128r^6+1728r^5+8968 r^4+21490r^3+20607 r^2-1228 r-8628}{(r+2)(r+3)(r+4)(2r-1)(2r+1)(2r+3)(2r+5)(2r+7)}+2H_{r}}\,,
\ee
and for the magnetic case:
\be
\label{eq:c_m}
\ba{rccll}
\ds i{\cal A}^{(m)}_c(\omega\,,\K) &=& (-i)^{r+1}&\ds
\pa{\frac{E^2 c_r^{(J)}}{8\Lambda^5}}\omega \epsilon_{ikl}J^{jki_1\cdots i_r}(\omega)&\\
&&\ds\int_{\pp,\Q}&\ds \frac {q_l q_{i_1}\cdots q_{i_r}}{\paq{\Q^2-\pa{\omega+i\ta}^2}\paq{\pp^2-\pa{\omega+i\ta}^2}
\pa{\pp-\Q}^2 \pa{\pp-\K}^2}&\\
&&\ds\times \ds\sigma^*_{ab}(\omega,\K)&\ds
\Bigg\{+\omega^4 \delta_{ia}\delta_{jb}&\{\sigma^2\}\\
&&&\ds\quad+\omega^2 [ (q-p)_j p_a\delta_{ib}-p_i p_a\delta_{jb}-q_j (q-p)_a\delta_{ib} ]&\{A\sigma\}\\
&&&\ds\quad -\frac{1}{c_d} [p_i (q-p)_j + p_j (q-p)_i]p_a p_b&\{\phi \sigma\}\\
&&&\ds\quad -\frac{1}{c_d}q_j p_i p_a p_b&\{A\phi\}\\
&&&\ds\quad -q_j p_a [(\pp\cdot\Q)\delta_{bi} -p_i q_b]&\{A^2\}\Bigg\}\\
&\simeq&\ds i \mathcal{A}_{r0}^{(m)}(\omega\,,\K)&\ds\pa{G_N E \omega}^2
\paq{-\frac{\pa{\omega+i\ta}^2}{\tilde{\mu}^2}}^{\epsilon}\paq{-\frac2{ \epsilon^2_{\rm IR}}+\frac{\alpha^{(m)}_c(r)}{\epsilon}}&\,,
\ea\nonumber\\
\label{eq:cc_m}
\alpha^{(m)}_c(r)\equiv 4\paq{\frac{(r+1)(8 r^3+ 64 r^2 + 165 r +141)}{(r+2)(r+3)(2r+3)(2r+5)(2r+7)}+H_{r+1}}\,.
\ee

While leaving the details of the computation to the Appendix, we highlight
that, contrarily to the single pole that contains both UV and IR divergences,
the double pole (due uniquely to the $\{\sigma^2\}$ contribution) is purely IR
and universal, as expected from the exponentiation of the simple tail IR
divergence.
Indeed, expanding the divergent phase at order $(G_NE\omega)^2$, one obtains schematically
\be
\label{eq:exp}
\ba{rcl}
\ds e^{iG_NE\omega\pa{\frac 2{\epsilon_{\rm IR}}-2\rho^{(e,m)}}}&\simeq&\ds
1+iG_NE\omega\pa{\frac 2{\epsilon_{\rm IR}}-2\rho^{(e,m)}}\\
&&\ds
-\pa{G_NE\omega}^2\pa{\frac 2{\epsilon^2_{\rm IR}}-\frac{4\rho^{(e,m)}}{\epsilon_{\rm IR}}+O(\epsilon^0_{\rm IR})}
+O\pa{(G_NE\omega)^3}\,;
\ea
\ee
i.e., the knowledge of the $O(\epsilon^0_{\rm IR})$ tail term, in Eq.~(\ref{eq:exp})
indicated generically with $\rho^{(e,m)}$ in the term linear in $G_NE\omega$, allows one to isolate the simple pole IR
divergence of the tail-of-tail process (quadratic piece in $G_NE\omega$), which, in turn, can be subtracted from Eqs.
(\ref{eq:c_e}) and (\ref{eq:c_m}) to finally identify the UV one.

\section{Summary and Discussion}
\label{sec:sdisc}
The general structure of the emission amplitude, including post-Minkowskian multipolar corrections, is
\be
\label{eq:em_rad}
\ba{rcl}
\ds i \mathcal{A}(\omega,\K)&=&\ds e^{i\frac{\phi_{\rm IR}(\omega)}{\epsilon_{\rm IR}}}\sum_r\frac{(-i)^{r+1}}{2\Lambda}\sigma_{ij}^{*}(\omega,\K) k_{i_1}\cdots k_{i_r}\\
&&\ds\qquad\times\left[{c_r}^{(I)} \omega^2 I^{iji_1\cdots i_r}_{rad}(\omega)+c_r^{(J)}\omega\epsilon_{ikl} k_l J^{jki_1 \cdots i_r}_{rad}(\omega)\right]\,,
\ea
\ee
where $(I,J)^{jki_1 \cdots i_r}_{rad}$ are the so-called \emph{radiative multipoles} and
\be
\phi_{\rm IR}(\omega)\equiv 2G_N E \omega\pa{\frac{\omega^2}{\tilde{\mu}^2}}^{\epsilon_{\rm IR}/2}
\ee
is the coefficient of the IR pole,
which is, however, unobservable, because it represents a global phase shift common
to every multipolar contribution of the emission amplitude. Likewise unobservable is the logarithmic term
generated in $\phi_{\rm IR}/\epsilon_{\rm IR}$ at $\epsilon_{\rm IR}^0$ order.

Differently from IR divergences, UV ones make their first appearance at
second PM order and have an important physical interpretation, as they signal
the breakdown of the point particle approximation for the composite object and
must be regularized.
Applying standard regularization and renormalization procedures,
one can obtain physical results from our UV-divergent amplitude.
Note that, while such procedures have been first developed and are routinely
used in \emph{quantum} field theory, they can be also applied here to our
completely classical setting, as they depend on the \emph{field theory}
nature of the problem.

The divergence can be absorbed in the definition of the
(divergent) \emph{bare} source multipoles $(I,J)^{iji_1\cdots i_r}_B$, related to
the renormalized, finite source multipoles $(I,J)^{iji_1\cdots i_r}_R$
by a divergent factor:
\be
\label{eq:bare}
I^{iji_1\cdots i_r}_B(\omega)=\paq{1-\frac{\beta^{(e)}(r)}{2\epsilon_{UV}}\pa{G_NE\omega}^2}I^{iji_1\cdots i_r}_R(\omega,\mu)
\ee
and analogously for the magnetic multipoles.
From the calculation of the previous section, we found
\begin{align}
\label{eq:beta_e}
\beta^{(e)}(r)&\equiv2\pa{\alpha^{(e)}_a+\alpha^{(e)}_b+\alpha^{(e)}_c-4\kappa_{r+2}}= -2\frac{15r^4+150 r^3+568r^2+965r+642}{(r +2)(r+3)(2r +3)(2r +5)(2r +7)}\,,\\
\beta^{(m)}(r)&\equiv2\pa{\alpha^{(m)}_a+\alpha^{(m)}_b+\alpha^{(m)}_c-4\pi_{r+2}}= -2\frac{15r^4+150 r^3+568r^2+965r+642}{(r +2)(r+3)(2r +3)(2r +5)(2r +7)}\,,\nn\\
\label{eq:beta_m}\end{align}
where the electric coefficients $\beta^{(e)}(r)$ have been first determined in
Ref. \cite{Blanchet:1987wq} and we have computed in this work for the first time the
expression for the magnetic ones $\beta^{(m)}(r)$, which turns to be equal to the electric case\footnote{Such equality have been independently established in \cite{Fucito:2024wlg}, and recently confirmed in \cite{Ivanov:2025ozg}.
The finding of \cite{Fucito:2024wlg} induced us to reconsider our result for
the magnetic case correcting a mistake in equation (\ref{eq:cc_m})
in the second
version of this work, where it was erroneously concluded that the electric
and magnetic beta functions are different.}; for this reason both functions will be simply denoted as $\beta(r)$ from now on.

Substituting for $(I,J)$ in the amplitudes of the previous section the bare
source multipoles $(I,J)_B$ expression (\ref{eq:bare}), one finds finite
expressions for the amplitudes in terms of the renormalized multipoles.
Hence, up to the second post-Minkowskian order,
\emph{radiative multipoles} entering the physical amplitude (\ref{eq:em_rad}) can
be related to renormalized source multipoles via
\be
\label{eq:Irr}
\ba{lcl}
\ds I^{iji_1\cdots i_r}_{rad}(\omega)&\simeq&\ds I_R^{iji_1\cdots i_r}(\omega,\mu)
e^{-2iG_NE\omega\kappa_{r+2}}\\
&&\ds\qquad\times\paq{1+\pi G_N |\omega|E+
\frac{\beta(r)}2\pa{G_N E\omega}^2\pa{\log\frac{\omega^2}{\tilde\mu^2}+O(\epsilon^0)}}
\ea
\ee
and analogously for the magnetic case.
In this renormalization procedure, which relies on large-scale physics and does not depend on the specific UV structure of the system,
the finite $O(\epsilon^0)$ contribution is left undetermined and must be
fixed by comparison with observations or a fine-grained description of the source.

The leading-order (real) tail correction $\pi G_N E|\omega|$ is multipole independent
and is generated by the imaginary part of the $\epsilon_{\rm IR}^{-1}\pa{-\pa{\omega+i\ta}^2}^{\epsilon_{\rm IR}}$ term, which is finite
for $\epsilon_{\rm IR}\to 0$, as derived in Sec. \ref{ssec:met_tail}.
At the same post-Minkowskian order of the leading tail, there are further finite contributions, not displayed in Eq.~(\ref{eq:Irr}),
coming from the angular momentum (failed) tail and the memory
effect, which for compact binaries are suppressed with respect to the leading order in the post-Newtonian expansion by a factor of $v^2$.
The expression of such terms in the time domain can be
found in Ref. \cite{Faye:2014fra} for the first multipoles ($r=0,1$). As to the finite phases proportional to
$\kappa_{r+2}$ and $\pi_{r+2}$, they are, in principle, observable as
discussed in the introduction, because they are not universal.

Note that, as the physical emission amplitude is directly related to the
radiative multipoles $(I,J)_{rad}$ and cannot depend on the arbitrary
renormalization scale $\mu$, the renormalized multipoles must acquire at
2PM order a $\mu$
dependence to compensate the explicit dependence on $\mu$ of the expression
(\ref{eq:Irr}), hence the argument $\mu$ added to $(I,J)_R$
already in Eq.~(\ref{eq:bare}).

This leads to the renormalization group equation
\be
\frac{{\rm d}I^{iji_1\cdots i_r}_R(\omega,\mu)}{{\rm d}\log\mu}=\beta(r) \pa{G_NE\omega}^2I^{iji_1\cdots i_r}_R(\omega,\mu)\,,
\ee
 which is solved by \cite{Goldberger:2009qd}
\be
I^{iji_1\cdots i_r}_R(\omega,\mu)=\pa{\frac{\mu}{\mu_0}}^{\beta(r)\pa{G_NE\omega}^2}I^{iji_1\cdots i_r}_R(\omega,\mu_0)
\ee
and analogously for the magnetic multipoles $J^{iji_1\cdots i_r}_R(\omega,\mu)$.

The above equations make manifest the role of $\beta(r)$  as beta
  functions controlling the running of the radiative multipoles. The
  renormalization group equation of the electric quadrupole \cite{Blanchet:2019rjs} has been used to resum an infinite series of leading logarithmic
  terms in the gauge-invariant expression for energy and angular momentum of
  compact binaries.
While the phenomenological impact for gravitational waveforms is expected
  to be modest (we remind that the lowest-order UV logarithms enter the
  waveform is 3PN), with the beta functions known at all multipole orders it is
  possible to compute the leading logarithmic terms in the energy, which are of
  the type $(M^2\log)^n\times \pa{d^{n+2}I_L/dt^{n+2}}^2$, at subleading PN orders. 
  This allows the possibility of additional, highly nontrivial checks with the
  PN-expanded version of extreme mass ratio results, in analogy to what is done
  in Ref. \cite{Blanchet:2019rjs} at leading PN order, where terms given in
  Ref. \cite{Kavanagh:2015lva} for $n\leq 7$ (contributing to the energy of circular
  orbit up to 22PN order) could be explicitly checked.

Knowledge of all the beta functions allows for an extension of this approach: In particular, before the present work, only the first magnetic coefficient $\beta^{(m)}(0)$ coefficient was known and found to be equal to the electric one; we have confirmed that this equality holds for all multipoles, in agreement with \cite{Fucito:2024wlg} and \cite{Ivanov:2025ozg}.
In the case of compact binaries, alternatively to the universal renormalization
procedure, one can exploit the explicit knowledge of the system at small scales,
as has been done in the 4PN study of the conservative sector \cite{Foffa:2019yfl},
to cancel the UV divergence from the multipolar dynamics (also called the far zone) with an IR divergence coming from the PN-expanded dynamics of individual
binary components interacting via the exchange of longitudinal gravitational modes (the near zone).
In this case, the cancellation should come from the explicit determination of
the \emph{source} multipoles in terms of the binary constituents' variables
at 3PN order, as preliminary confirmed by Ref. \cite{Luc},
and the previously undetermined $O(\epsilon^0)$ term appearing in Eq.
(\ref{eq:Irr}) is expected to be unambiguously predicted in terms of the UV
details of the system.

\section*{Acknowledgements}
We would like to thank the authors of \cite{Fucito:2024wlg} for informing us about their findings.
The work of R.S. is partly supported by CNPq by Grant No. 312320/2018-3.
R. S. thanks ICTP-SAIFR FAPESP Grant No. 2016/01343-7.
The work of G. L. A. is financed in part by the Coordena\c{c}\~{a}o de Aperfei\c{c}oamento de Pessoal de N\'{i}vel Superior - Brasil (CAPES) - Finance Code 001.
S.F. is supported by the Fonds National Suisse and by the SwissMap NCCR.

\appendix{}
\section{Useful integrals}
\label{sec:app}
All integrals involved in tail diagram computations, as well as in amplitudes
(a) and (b) of the tail-of-tail process,
can be derived (eventually after iteration) from the following standard one-loop
scalar master integrals:
\be
\label{eq:app_std}
\ba{l}
\ds J_{ab}(\Q) \equiv \ds\int_\Pp\frac 1{\Pp^{2a}\pa{\Pp-\Q}^{2b}}=\ds(\Q^2)^{d/2-a-b}\frac{\Gamma(a+b-d/2)\Gamma(d/2-a)\Gamma(d/2-b)}{(4\pi)^{d/2}\Gamma(a)\Gamma(b)\Gamma(d-a-b)}\,,\\
\ds I_a(\omega)\equiv \int_\Q\frac 1{\paq{\pa{\K-\Q}^2}^a[\Q^2-\pa{\omega+i\ta}^2]}
=\paq{-\pa{\omega+i\ta}^2}^{d/2-a-1}\dfrac{\Gamma(a+1-d/2)\Gamma(d-2a-1)}
{\pa{4\pi}^{d/2}\Gamma(d-a-1)}\,,
\ea
\ee
where in the $I_a$ equation it is understood that $\K^2=\omega^2$.
The eventual presence of tensorial structures at the numerator is accounted by
the usual scalarization procedure plus some combinatorics.
For instance, borrowing notation from Ref. \cite{Smirnov:2004ym},
  \be
  \label{eq:Strick}
  \ba{l}
  \ds\int_\Q\frac{q_{i_1}\dots q_{i_n}}{\paq{\pa{\K-\Q}^2}^a\paq{\Q^2-(\omega+i\ta)^2}^b}=
  \sum_{m=0}^{[n/2]}S_{a,b}(n,m)\,,\\
  \ds S_{a,b}(n,m)\equiv
  \frac{\paq{-\pa{\omega+i\ta}^2}^{d/2-a-b+m}}{2^m\pa{4\pi}^{d/2}}
  \frac{\Gamma(a+b-d/2-m)\Gamma(a+n-2m)\Gamma(d+2m-2a-b)}{\Gamma(a)\Gamma(b)\Gamma(d+n-a-b)}\\
  \qquad\times \{[\delta]^m[k]^{n-2m}\}_{i_1\cdots i_n}\,,  
  \ea
  \ee
    where $\{[\delta]^m[k]^{n-2m}\}_{i_1\cdots i_{n}}$ is symmetric in its $n$ indices,
  it involves $m$ Kronecker deltas and $n-2m$ occurrences of $k$ vectors, and $[n/2]$ is the integer part of $n/2$.
To write the amplitude of the diagrams in Figs. 3(a) and 3(b)
we preliminarily define
\begin{align}
\tilde{\delta}_{abcd}&\equiv \delta_{ac}\delta_{bd}+\delta_{ad}\delta_{bc}-\frac{2}{d-2}\delta_{ab}\delta_{cd}\,,\\
D^{(1)}_{abcd} &= \delta_{ab}\delta_{cd}-\frac 12 \delta_{ac}\delta_{bd}\,,\\
D^{(2)}_{abcdef} &= \frac 14 \delta_{ab}\delta_{ce}\delta_{df}+\frac 12 \delta_{cd}\delta_{ae}\delta_{bf}-\delta_{ac}\delta_{be}\delta_{df}\,,
\\
D^{(3)}_{abcdefmn} &= -\frac 14 \delta_{ab}\delta_{mn}\delta_{ce}\delta_{df}-\frac{1}{2}\delta_{mn}\delta_{cd}\delta_{ae}\delta_{bf} + \delta_{mn}\delta_{ac}\delta_{be}\delta_{df} + \frac 12 \delta_{am}\delta_{bn}\delta_{ce}\delta_{df}
\nonumber\\
&+D^{(1)}_{arbs}\times
(\delta_{rc}\delta_{se}\delta_{mf}\delta_{nd}-\delta_{rc}\delta_{se}\delta_{md}\delta_{nf}
+\delta_{rm}\delta_{se}\delta_{cd}\delta_{nf}-\delta_{rc}\delta_{sm}\delta_{ef}\delta_{nd})\,,\\
D^{(4)}_{abcdef} & = 2 \delta_{ae}\delta_{bc}\delta_{df} -\delta_{ad}\delta_{be}\delta_{cf} -2\delta_{af}\delta_{be}\delta_{cd} +\delta_{af}\delta_{bc}\delta_{de}\,.
\end{align}
In the diagram in Fig. 3(a) the propagator labeled by $q$ can carry a $\sigma$ or an $A$
polarization (the others are fixed, as only $\phi$ couples to the conserved energy $E$), the two separate contributions being
\begin{align}
i \mathcal{A}^{(e)}_{a\,,\sigma}(\omega\,,\K) &= \frac{(-i)^{r+1}E^2 c_{r}^{(I)}}{16 c_d\Lambda^5} \omega^2 I^{iji_1\cdots i_r}(\omega)
\int_{\Q}\frac {q_{i_1}\cdots q_{i_r}}{\pa{\K-\Q}^2[\Q^2-\pa{\omega+i\ta}^2]}
\int_{\pp}\frac 1{\pa{\pp-\Q}^2\pa{\K-\pp}^2} \nonumber \\
&\quad \times \pa{p-q}_{\beta}(p-k)_{\delta}D^{(1)}_{\alpha\beta\gamma\delta}\nonumber\\
&\quad\times
\Bigg\{ D^{(2)}_{abcdef}\omega^2 \tilde{\delta}_{ab\alpha\gamma}
[\delta_{ic}\delta_{jd}\sigma^{*}_{ef}(\omega\,,\K)+\delta_{ie}\delta_{jf}\sigma^{*}_{cd}(\omega\,,\K)] \\
&\qquad +D^{(3)}_{abcdefmn}\bigg[-\delta_{ia}\delta_{jb}(\pa{q-k}_m k_n \tilde{\delta}_{cd\alpha\gamma}\sigma^{*}_{ef}(\omega\,,\K)+k_m\pa{q-k}_n  \tilde{\delta}_{ef\alpha\gamma}\sigma^{*}_{cd}(\omega\,,\K))    \nonumber\\
&\qquad\qquad\qquad\qquad\qquad+\delta_{ic}\delta_{jd}q_m (\tilde{\delta}_{ab\alpha\gamma}k_n\sigma^{*}_{ef}(\omega\,,\K)+\tilde{\delta}_{ef\alpha\gamma}\pa{q-k}_n\sigma^{*}_{ab}(\omega\,,\K))      \nonumber\\
&\qquad\qquad\qquad\qquad\qquad+\delta_{ie}\delta_{jf}q_n (\tilde{\delta}_{ab\alpha\gamma}k_m\sigma^{*}_{cd}(\omega\,,\K)+\tilde{\delta}_{cd\alpha\gamma}\pa{q-k}_m\sigma^{*}_{ab}(\omega\,,\K))
\bigg]
\Bigg\}\nn
\end{align} 
and
\begin{align}
  i \mathcal{A}^{(e)}_{a\,,A}(\omega\,,\K) &= \frac{-(-i)^{r+1}E^2 c_{r}^{(I)}}{32 c_d\Lambda^5} \omega^2\! I^{iji_1\cdots i_r}(\omega)
\int_{\Q}\frac {q_{i_1}\cdots q_{i_r}}{\pa{\K-\Q}^2[\Q^2-\pa{\omega+i\ta}^2]}\int_{\pp}\frac 1{(\pp-\Q)^2\pa{\K-\pp}^2} \nonumber \\
&\quad \times
D^{(1)}_{\alpha\beta\gamma\delta} D^{(4)}_{abcdei}
q_j [\pa{p-q}_{\beta}(p-k)_{\delta}+(p-q)_{\delta}(p-k)_{\beta}] (q-k)_c \tilde{\delta}_{\alpha\gamma de}
\sigma^{*}_{ab}(\omega\,,\K)\,.
\end{align}
Similarly, for the magnetic case
\begin{align}
  i \mathcal{A}^{(m)}_{a\,,\sigma}(\omega\,,\K) &= \frac{(-i)^{r+1}E^2 c_{r}^{(J)}}{64 c_d\Lambda^5} \omega\epsilon_{ikl} J^{jki_1\cdots i_r}(\omega)
\int_{\Q}\frac {q_{i_1}\cdots q_{i_r}}{(\K-\Q)^2[\Q^2-\pa{\omega+i\ta}^2]}\int_{\pp}\frac 1{\pa{\pp-\Q}^2(\K-\pp)^2} \nonumber \\
&\quad \times   D^{(1)}_{\alpha\beta\gamma\delta}q_l [(p-q)_{\beta}(p-k)_{\delta}+(p-q)_{\delta}(p-k)_{\beta}]\nonumber\\
&\qquad\times\Bigg\{ D^{(2)}_{abcdef}\omega^2 \tilde{\delta}_{ab\alpha\gamma}
[\tilde{\delta}_{ijcd}\sigma^{*}_{ef}(\omega\,,\K)+\tilde{\delta}_{ijef}\sigma^{*}_{cd}(\omega\,,\K)] \nonumber \\
&\qquad \qquad+D^{(3)}_{abcdefmn}\bigg[-\tilde{\delta}_{ijab}((q-k)_m k_n \tilde{\delta}_{cd\alpha\gamma}\sigma^{*}_{ef}(\omega\,,\K)+k_m(q-k)_n  \tilde{\delta}_{ef\alpha\gamma}\sigma^{*}_{cd}(\omega\,,\K))    \nonumber\\
  &\qquad\qquad\qquad\qquad\qquad+\tilde{\delta}_{ijcd}q_m (\tilde{\delta}_{ab\alpha\gamma}k_n\sigma^{*}_{ef}(\omega\,,\K)+\tilde{\delta}_{ef\alpha\gamma}(q-k)_n\sigma^{*}_{ab}(\omega\,,\K))\nn\\
  &\qquad\qquad\qquad\qquad\qquad+\tilde{\delta}_{ijef}q_n (\tilde{\delta}_{ab\alpha\gamma}k_m\sigma^{*}_{cd}(\omega\,,\K)+\tilde{\delta}_{cd\alpha\gamma}(q-k)_m\sigma^{*}_{ab}(\omega\,,\K))
\bigg]\Bigg\}
\end{align}
and 
\begin{align}
  i \mathcal{A}^{(m)}_{a\,,A}(\omega\,,\K)\! &=\! \frac{-(-i)^{r+1}E^2 c_{r}^{(J)}}{64 c_d\Lambda^5} \omega\epsilon_{ikl} J^{jki_1\cdots i_r}(\omega)
\int_{\Q}\!\frac {q_{i_1}\cdots q_{i_r}}{\pa{\K-\Q}^2\![\Q^2-\pa{\omega+i\ta}^2]}\!\!\int_{\pp}\!\frac 1{\pa{\pp-\Q}^2\!(\K-\pp)^2} \nonumber \\
&\quad \times D^{(1)}_{\alpha\beta\gamma\delta}\! D^{(4)}_{abcdei}
q_lq_j [\pa{p-q}_{\beta}(p-k)_{\delta}+\pa{p-q}_{\delta}(p-k)_{\beta}](q-k)_c \tilde{\delta}_{\alpha\gamma de}
\sigma^{*}_{ab}(\omega\,,\K)\,.
\end{align}
The calculation of the diagram in Fig. 3(b) is similar and gives
\begin{align}
i{\cal A}^{(e)}_b(\omega\,,\K) &= (-i)^{r+1} \left(\frac{E^2 c_r^{(I)}}{16\Lambda^5}\right)I^{iji_1\cdots i_r}(\omega)
\int_{\Q}\frac {q_{i_1}\cdots q_{i_r}}{[\Q^2-\pa{\omega+i\ta}^2]}
\int_{\pp}\frac 1{\pa{\pp-\Q}^2\pa{\K-\pp}^2}
\nonumber\\
&\Bigg\{\delta_{ib}\omega^4+\delta_{ib}\frac {\omega^2}{c_d}\pa{\pp-\Q}\cdot(\pp-\K) 
-\frac {4}{c_d}\omega^2 (p-q)_i(p-k)_b\Bigg\}\sigma_{bj}^{*}(\omega\,,\K)\,,
\end{align} 
\begin{align}
&i{\cal A}^{(m)}_b(\omega\,,\K) = (-i)^{r+1} \left(\frac{E^2 c_r^{(J)}}{16\Lambda^5}\right)\omega\epsilon_{ikl}J^{jki_1\cdots i_r}(\omega)
\int_{\Q}\frac {q_{i_1}\cdots q_{i_r}}{[\Q^2-\pa{\omega+i\ta}^2]}
\int_{\bf p}\frac 1{\pa{\pp-\Q}^2(\K-\pp)^2}
\nonumber\\
&\qquad \times q_l\Bigg\{\omega^2\delta_{ai}\delta_{bj}+\frac1{c_d}\pa{\pp-\Q}\cdot(\pp-\K)\delta_{ai}\delta_{bj}
-\frac2{c_d}(p-q)_a \left[(p-k)_i \delta_{bj}+(p-k)_j \delta_{bi}\right] \Bigg\}\sigma_{ab}^{*}(\omega\,,\K)\,.
\end{align}

Tail-of-tail amplitude (c) is more complicated, as it involves the following
family of two-loop integrals (always $\omega^2=\K^2$):
\be
I_{in}\paq{a_1,a_2,a_3,a_4,a_5}\equiv\int_{\pp,\Q}
\frac 1{\paq{\Q^2-\pa{\omega+i\ta}^2}^{a_1}\paq{\pp^2-\pa{\omega+i\ta}^2}^{a_2}\pa{\pp-\Q}^{2a_3} \pa{\pp-\K}^{2a_4}\pa{\Q-\K}^{2a_5}}\,.\nn\\
\ee
The general expression is long and complicated; here, we focus only on the  part which is singular in the $d\rightarrow 3$ limit, which is the relevant one in the renormalization procedure.
Using the standard technique of integration by parts implemented by the software Reduze \cite{vonManteuffel:2012np}, one can express the main scalar integral as
\be
\ba{rcl}
\ds I_{in}\paq{1,1,1,1,0}&=&\ds
\frac 1{\omega^2}\frac{3d-8}{4\pa{d-3}}\int_{\pp,\Q}
\frac 1{\pa{\pp^2-\pa{\omega+i\ta}^2}\pa{\pp-\Q}^2\pa{\pp-\K}^2}+\\
&&\ds\frac 1{4\omega^4}\frac{d^2+4d-4}{\pa{d-3}^2}
\int_{\pp,\Q}\frac 1{\pa{\Q^2-(\omega+i\ta)^2}\pa{\pp^2-(\omega+i\ta)^2}}\,.
\ea
\ee
When reducing tensor integral to scalar ones, the following results are needed, for $m\,,n\in\mathbb{N}$:
\be
\ba{rcl}
\ds I_{in}\paq{1,1,1,1,0}&\simeq& \ds-\paq{128 \pi^2 \pa{\omega+i\ta}^2\epsilon^2}^{-1}+O(\epsilon^0)\\
\ds I_{in}\paq{1,1,1,1,-n}&\simeq&\ds -\frac{\paq{4\pa{\omega+i\ta}^2}^n}{n}
  \frac{1}{128 \pi^2 \pa{\omega+i\ta}^2\epsilon}+O(\epsilon^0)\,\quad {\rm for}\ n\geq 1\\
I_{in}\paq{1, -m, 1, 1, -n}&\simeq&\ds
-\frac{(-1)^m \paq{\pa{\omega+i\ta}^2}^{m+n} \Gamma (m+2 n+1)}{64 \pi ^{3/2} \epsilon  \Gamma (n+1) \Gamma \left(m+n+\frac{3}{2}\right)}+O(\epsilon^0)\\
\ds I_{in}\paq{1,1,-m,1,-n}&\simeq&\ds -\frac{\paq{4\pa{\omega+i\ta}^2}^{m+n}}{32\pi^2\pa{m+n+1}\epsilon}+O(\epsilon^0)\\
\ds I_{in}\paq{1,1,1,-m,-n}&\simeq &\ds I_{in}\paq{1,1,-m,1,-n}+O(\epsilon^0)\\
\ds I_{in}\paq{1,1,0,0,-n}&\simeq &\ds O(\epsilon^0)\,.
\ea
\ee
From there, one can compute the only unknown parameter involved in the following equation:
\be
\label{eq:app}
\int_{\pp,\Q}\frac {q_{(i_1\cdots i_r)}}{{\cal D}_{(tail)^2}}\equiv\int_{\pp,\Q}\frac {q_{(i_1\cdots i_r)}}{\paq{\Q^2-\pa{\omega+i\ta}^2}\paq{\pp^2-\pa{\omega+i\ta}^2}\pa{\pp-\Q}^2 \pa{\pp-\K}^2}\simeq\frac{{\cal A}_r}{\omega^2} k_{(i_1\cdots i_r)}\,,
\ee
$k_{(i_1\cdots i_r)}$ being (still following the notation of Ref. \cite{Smirnov:2004ym}) the symmetric traceless (STF) combination  of $k^i$'s.
In detail:
\be
\ds\frac{{\cal A}_r}{\omega^2} k_{(i_1\cdots i_r)}\times k_{i_1}\cdots k_{i_r}&=&\ds {\cal A}_r C_r\pa{\omega^2}^{r-1}=\int_{\pp, \Q}\frac {q_{(i_1\cdots i_r)}k_{i_1}\cdots k_{i_r}}{{\cal D}_{(tail)^2}}\nn\\
&=&\ds \sum_{j=0}^{\paq{\frac r2}}b_{r,j}\pa{\omega^2}^j\int_{\pp,\Q}\frac{\pa{\Q^2}^j\pa{{\Q}\cdot\K}^{r-2j}}{{\cal D}_{(tail)^2}}\nn\\
&=&\ds\pa{\omega^2}^r \sum_{j=0}^{\paq{\frac r2}}\sum_{a_1=0}^{r-2j}b_{r,j}\pa{-2\omega^2}^{-a_1}\binom{r-2j}{a_1}I_{in}\paq{1,1,1,1,-a_1}\nn\\
&\simeq&\ds -\frac{\pa{\omega^2}^{r-1}C_r}{128 \pi^2}\paq{\frac1{\epsilon^2}-\frac{2 H_r}{\epsilon}}\,,
\ee
with $H_r$ the harmonic number and 
\be
b_{r,j} \equiv \frac{r!}{4^j j! (r-2j)! (2-r-d/2)_j}\,,\quad C_r\equiv \sum_{i=0}^{\paq{\frac r2}}b_{r,i}=\frac{\Gamma\pa{d+r-2}}{\pa{d-3}!!\pa{d+2r-4}!!}\,,\nn
\ee
$(a)_b$ being the Pochhammer symbol.
Moreover, one needs to compute
  integrals like the one above, with the addition of up to four $p_i$'s and up to two extra $q_j$'s (not involved in the STF combination with the other $q^{i_r}$'s), and this can be achieved
via a tedious but straightforward scalarization procedure.

For instance, for one extra $p_i$, one can write
\be
&&\int_{\pp, \Q}\frac {q_{(i_1\cdots i_r)}p_i}{{\cal D}_{(tail)^2}}=\frac{{\cal A}^{(p)}_r}{\omega^2}k_{(i_1\cdots i_r)k_i}+{\cal B}^{(p)}_r \delta_{i(i_1}k_{i_2\cdots i_r)}\,,
\ee
and two independent contractions are needed to solve the linear system. One is the same as above, while another can be obtained by contracting the index $i$ with one of the STF indices. The integrals are just slightly more complicated with
respect to the one needed in Eq. (\ref{eq:app}). Adding extra factors to the
integrand does not introduce insurmountable complications.

For one extra, non-STF, $q$ factor, one can proceed in the same way:
\be
&&\int_{\pp, \Q}\frac {q_{(i_1\cdots i_r)}q_j}{{\cal D}_{(tail)^2}}=\frac{{\cal A}^{(q)}_r}{\omega^2}k_{(i_1\cdots i_r)k_j}+{\cal B}^{(q)}_r \delta_{j(i_1}k_{i_2\cdots i_r)}\,,
\ee
and solve the associated linear system. Actually, by noticing that
\be
q_{(i_1\cdots i_r)}q_j=q_{(i_1\cdots i_rj)}+\frac r{d+2r -2}\Q^2\delta_{j(i_1}q_{i_2\cdots i_r)}
\ee
one can straightforwardly derive
\be
\int_{\pp, \Q}\frac {q_{(i_1\cdots i_r)}q_j}{{\cal D}_{(tail)^2}}&=&\int_{\pp, \Q}\frac {q_{(i_1\cdots i_rj)}}{{\cal D}_{(tail)^2}}+\frac r{d+2r -2}\int_{\pp, \Q}\Q^2\frac{\delta_{j(i_1}q_{i_2\cdots i_r)}}{{\cal D}_{(tail)^2}}\nn\\
&=&\frac{{\cal A}_{r+1}}{\omega^2}k_{(i_1\cdots i_r j)} +\frac r{d+2r -2}{\cal A}_{r-1} \delta_{j(i_1}k_{i_2\cdots i_r)}\\
&=&\frac{{\cal A}_{r+1}}{\omega^2}k_{(i_1\cdots i_r) j} +\frac r{d+2r -2}\paq{{\cal A}_{r-1}-{\cal A}_{r+1}} \delta_{j(i_1}k_{i_2\cdots i_r)}\nn\,.
\ee

\end{document}